\documentclass[12pt,preprint]{aastex}
\usepackage{amsmath}

\shortauthors{Wang, Ulrich, \& Harvey}
\shorttitle{Magnetograph Saturation and the Open Flux Problem}

\begin{document}

\title{Magnetograph Saturation and the Open Flux Problem}
\author{Y.-M. Wang}
\affil{Space Science Division, Naval Research Laboratory, Washington, 
DC 20375, USA}
\email{yi.wang@nrl.navy.mil}
\author{R. K. Ulrich}
\affil{Department of Physics and Astronomy, University of California, 
Los Angeles, CA 90095, USA}
\email{ulrich@astro.ucla.edu}
\and
\author{J. W. Harvey\altaffilmark{1}}
\affil{National Solar Observatory, Boulder, CO 80303, USA}
\email{jharvey@nso.edu}
\altaffiltext{1}{Also at Lunar and Planetary Laboratory, University of Arizona, Tucson, AZ 85719, USA.}

\begin{abstract}
Extrapolations of line-of-sight photospheric field measurements predict 
radial interplanetary magnetic field (IMF) strengths that are factors of 
$\sim$2--4 too low.  To address this ``open flux problem,'' we reanalyze 
the magnetograph measurements from different observatories, with 
particular focus on those made in the saturation-prone Fe I 525.0 nm line 
by the Mount Wilson Observatory (MWO) and the Wilcox Solar Observatory (WSO).  
The total dipole strengths, which determine the total open flux, 
generally show large variations among observatories, even when their 
total photospheric fluxes are in agreement.  However, the MWO and WSO 
dipole strengths, as well as their total fluxes, agree remarkably well 
with each other, suggesting that the two data sets require the same 
scaling factor.  As shown earlier by Ulrich et al., the saturation 
correction $\delta^{-1}$ derived by comparing MWO measurements in the 
525.0 nm line with those in the nonsaturating Fe I 523.3 nm line depends 
sensitively on where along the irregularly shaped 523.3 nm line wings 
the exit slits are placed.  If the slits are positioned so that the 523.3 
and 525.0 nm signals originate from the same height, $\delta^{-1}\sim 4.5$ 
at disk center, falling to $\sim$2 near the limb.  When this correction 
is applied to either the MWO or WSO maps, the derived open fluxes are 
consistent with the observed IMF magnitude.  Other investigators 
obtained scaling factors only one-half as large because they sampled 
the 523.3 nm line farther out in the wings, where the shift between the 
right- and left-circularly polarized components is substantially smaller.
\end{abstract}

\section{Introduction}
When extrapolated into the heliosphere, solar magnetograph measurements 
underestimate the radial interplanetary magnetic field (IMF) strength by 
factors of two or more (see, e.g., Wang \& Sheeley 1988, 1995; 
Riley et al. 2014, 2019; Jian et al. 2015; Linker et al. 2017; 
Wallace et al. 2019; Badman et al. 2021).  Two possible reasons for this 
large discrepancy (called the ``open flux problem'' by Linker et al. 2017) 
are that the magnetographs may be greatly underestimating the amount 
of large-scale flux threading the photosphere, or that the total amount 
of open magnetic flux on the Sun is much greater than predicted by the 
extrapolation models.

The most widely used technique for extrapolating line-of-sight measurements 
of the photospheric field into the corona and heliosphere is the 
potential-field source-surface (PFSS) method (Schatten et al. 1969; 
Altschuler \& Newkirk 1969).  In this idealized model, the magnetic field 
satisfies the current-free condition $\nabla\times\mbox{\boldmath{$B$}} = 0$ 
out to a spherical ``source surface'' at heliocentric distance 
$r = R_{\rm ss}$, where the tangential field components are set to zero.  
At the inner boundary $r = R_\odot$, $B_r$ is matched to the observed 
photospheric field, which is deprojected by assuming it to be radial 
at the depth (below the temperature minimum) where it is measured 
(see Wang \& Sheeley 1992).  Contrary to the prescription of 
Altschuler \& Newkirk (1969), the line-of-sight components are not matched 
because the photospheric field is nonpotential; instead, $B_r$ is taken 
to be conserved across the narrow ``boundary layer'' where the field 
makes the transition from radial/nonpotential to nonradial/potential.  
This assumption may break down if the magnetograph aperture size 
is much less than a supergranular radius ($\sim$20$^{\prime\prime}$), 
since some of the photospheric flux may then escape sideways 
as it fans out with height; however, continuity can be restored 
through spatial averaging, which will not affect the large-scale field 
that is the concern of this paper. 

The total open flux is given by integrating the unsigned radial field 
over the source surface:
\begin{equation}
\Phi_{\rm open}(t) = R_{\rm ss}^2\int\vert B_r(R_{\rm ss},L,\phi,t)\vert
d\Omega,
\end{equation}
where $t$ denotes time, $L$ heliographic latitude, $\phi$ Carrington 
longitude, and $\Omega$ solid angle.  Since the open flux is distributed 
isotropically at 1 AU according to {\it Ulysses} magnetometer observations 
(Balogh et al. 1995; Smith \& Balogh 2008), the radial field strength 
at Earth ($r = r_{\rm E}$) is related to $\Phi_{\rm open}$ by
\begin{equation}
B_{\rm E}(t) = \frac{\Phi_{\rm open}(t)}{4\pi r_{\rm E}^2}.
\end{equation}
The isotropization of the flux, which occurs by $r\sim 10$--15 $R_\odot$ 
(see, e.g., Wang 1996; Zhao \& Hoeksema 2010; Cohen 2015), is due to the 
heliospheric sheet currents, which are not included in the PFSS model.

For the source surface radius $R_{\rm ss}$, the only free parameter 
in the PFSS model, a value of $2.5\pm 0.25$ $R_\odot$ has been shown to 
approximately reproduce the IMF sector structure during 1976--1982 
(Hoeksema 1984) and the configuration of \ion{He}{1} 1083.0 nm 
coronal holes during 1976--1995 (see Figure 2 in Wang et al. 1996).   
If such observational constraints are ignored, an obvious approach to 
reconciling the photospheric measurements with the observed radial IMF 
strength is simply to allow $R_{\rm ss}$ to take on much smaller values 
and to vary with time (see, e.g., Wang \& Sheeley 1988; Lee et al. 2011; 
Linker et al. 2017; Bale et al. 2019; Badman et al. 2020).  
Virtanen et al. (2020) fitted the near-Earth IMF variation during 
1967--2017 by varying $R_{\rm ss}$ between $\sim$3 and $\sim$1.5 $R_\odot$, 
and deduced that the solar corona abruptly shrank by more than 
a factor of two after the late 1990s.  However, they provided 
no independent observational evidence to support this conclusion, 
other than the overall decline in sunspot activity since 2000.  If the 
corona underwent such a contraction, the heliospheric current/plasma sheet 
extensions of helmet streamers would presumably also have moved inward, 
but SOHO/LASCO C2 images show little long-term change in the 
quasi-equilibrium positions of helmet streamer cusps, which continue 
to extend out to at least $r\sim 2.5$ $R_\odot$ as they did in 
1996.\footnote{Daily LASCO C2 movies from 1996 to the present may be viewed 
at \url{http://spaceweather.gmu.edu/seeds}.}  Likewise, a long-term 
factor-of-two decrease in $R_{\rm ss}$ should be accompanied by a 
corresponding systematic increase in the areas of coronal holes, which 
(to our knowledge) has not been observed.

Comparisons between the PFSS model and extrapolations based on the 
magnetohydrodynamical (MHD) equations show generally good agreement 
in the predicted interplanetary sector structure and in the configuration 
of coronal holes (Neugebauer et al. 1998; Riley et al. 2006; Cohen 2015).  
The MHD models themselves depend on assumptions about the nature and 
global distribution of coronal heating, which effectively constitute 
a set of free parameters that replaces the source surface radius.

It has been suggested that the actual amount of open flux on the Sun 
may be much greater than that associated with visible coronal holes, 
either because the hole boundaries are not well defined or because 
small pockets of open flux are ubiquitous within closed regions.  
However, recent studies employing a wide variety of coronal hole detection 
techniques (Linker et al. 2021; Reiss et al. 2021) show that uncertainties 
in the locations of the hole boundaries cannot account for the 
large difference between the total open flux in coronal holes and the 
observed IMF strength; the discrepancy remains a factor of $\sim$2--4.  
Some of the disagreement in the inferred areal sizes of extreme-ultraviolet 
coronal holes could be due to the fanning-out of the hole boundaries 
with height, so that the line of sight traverses both the dark coronal hole 
and the brighter loop material underneath; since the amount of open flux 
is determined by the boundary at the photospheric level, the algorithms 
that predict the largest hole areas may be setting their intensity thresholds 
too high.\footnote{This problem would not affect coronal hole boundaries 
determined using the chromospheric \ion{He}{1} 1083.0 nm line.}

According to Fisk (2005), open flux is present outside coronal holes 
and is transported over the solar surface by undergoing interchange 
reconnection with closed loops.  There has been no direct observational 
evidence for such a global random-walk process, which would require 
the complete breakdown of the current-free approximation in the 
quiet corona.  The most likely site for continual interchange reconnection 
outside coronal holes is at their interface with the adjacent streamers; 
this may give rise to the raylike structure of the heliospheric 
plasma sheet that extends outward from the streamer cusps, 
but does not act to increase the total amount of open flux.

Riley et al. (2019) have speculated that most of the Sun's polar flux 
may be hidden from view, but this seems unlikely (even allowing for 
instrumental noise) given that each pole is tilted toward Earth by 
7$^\circ$ once a year.  The polar contribution to the open flux could be 
significantly underestimated if the field lines above latitude 70$^\circ$ 
have a poleward tilt of $\sim$6$^\circ$ at the photosphere, as suggested by 
Ulrich \& Tran (2013).  However, this effect would not be enough to double 
the open flux; nor would the many uncertainties associated with the 
construction of photospheric field maps, including the filling-in of 
missing polar data and the nonsimultaneous nature of the observations 
at different longitudes (see, e.g., Linker et al. 2017).

It has been proposed that the solar cycle variation of the near-Earth 
IMF strength is the result of interplanetary coronal mass ejections 
(see, e.g., Owens \& Crooker 2006; Schwadron et al. 2010; Owens et al. 2011).  
However, ICMEs cannot account for the large post-maximum peaks in $B_{\rm E}$ 
observed during 1982, 1991, 2002--2003, and 2014--2015, which reflected 
increases in the Sun's equatorial dipole strength associated with 
active longitudes or the emergence of large active-region complexes.  
In addition, using the ICME catalog of Richardson \& Cane (2010), 
Wang \& Sheeley (2015) showed that ICMEs contributed an average of 
only $\sim$20\% to $B_{\rm E}$ during the maxima of cycles 23 and 24, 
consistent with the study of Richardson \& Cane (2012).  Riley (2007) 
derived a much larger contribution by setting the average radial field 
strength of an ICME to 8 nT, more than twice the observed value.

Recent measurements with the FIELDS instrument on the Parker Solar Probe 
show that $B_rr^2$ is approximately conserved from 1 to 0.13 au, 
so that the discrepancy between magnetograph extrapolations and the 
observed IMF strength persists very close to the Sun (Badman et al. 2021).  
This also suggests that the problem is not caused by the increasing 
prevalence of disconnected flux or magnetic switchbacks at greater 
heliocentric distances.

From this discussion, we are led to conclude that the solution to 
the open flux problem lies not in the extrapolation methods, in 
overlooked sources of open flux on the Sun, or in the topological properties 
of the IMF, but most likely in the magnetograph measurements themselves.  
This is hardly surprising, given the many uncertainties involved 
in interpreting these measurements, such as saturation effects 
when the Zeeman shift becomes comparable to the line width 
(e.g., Howard \& Stenflo 1972; Frazier \& Stenflo 1972; Ulrich 1992; 
Ulrich et al. 2009; Demidov \& Balthasar 2009), weakening of the 
absorption lines due to the higher temperatures in magnetic regions 
(Chapman \& Sheeley 1968; Harvey \& Livingston 1969; 
Hirzberger \& Wiehr 2005), the tendency for the magnetic flux to be 
concentrated in the dark intergranular lanes rather than the 
bright granular cell centers (Plowman \& Berger 2020a,b,c), the fanning-out 
and weakening of the field with height, and the resulting dependence 
of the different effects on wavelength position relative to line center 
and on the center-to-limb angle, with horizontal field components 
or ``fringing'' becoming increasingly prevalent toward the limb.

In Section 2, we compare PFSS extrapolations of photospheric field maps 
from a variety ground- and space-based observatories/instruments, 
showing that the large differences in the predicted radial IMF strengths 
reflect differences in the measured dipole strengths, not in the 
photospheric fluxes.  Section 3 focuses on the question of the 
saturation correction required for the \ion{Fe}{1} 525.0 nm line 
used by the Mount Wilson Observatory (MWO) and the Wilcox Solar Observatory 
(WSO).  The corrected MWO and WSO open fluxes are compared with the 
observed IMF variation in Section 4, where the contribution of ICMEs 
is also discussed.  Our conclusions are summarized in Section 5.

\section{Deriving the Open Flux Using Photospheric Field Maps from 
Different Observatories}
As listed in Table 1, we employ synoptic maps of the photospheric magnetic 
field provided by MWO (1967--2013), WSO (1976--2021), National Solar 
Observatory (NSO) KPVT/SPM (1992--2003), SOHO/MDI (1996--2010), 
NSO/SOLIS/VSM (2003--2017), NSO/GONG (2006--2021), SDO/HMI (2010--2021), 
and Kislovodsk/STOP (2014--2021).  No saturation corrections were applied 
to the data after downloading them from the observatory 
websites.\footnote{The Level 1.8.2 MDI synoptic maps used here 
include a scaling factor of the order of 1.7, derived by Tran et al. (2005) 
from a comparison with calibrated MWO \ion{Fe}{1} 525.0 nm magnetograms.}  
In all cases, the photospheric field is assumed to be radially oriented 
and given by $B_r = B_{\rm los}/\cos{L}$, where the original 
line-of-sight measurements were taken around central meridian over 
a 27.3 day Carrington rotation (CR).  The low-resolution MWO and WSO maps 
were interpolated to 5$^\circ$ pixels in longitude and latitude, 
while the higher resolution maps from the other observatories 
were converted to a pixel size of 1$^\circ$.

Both MWO and WSO employ the \ion{Fe}{1} 525.0 nm line, which has a 
Land{\'e} factor $g = 3.0$ and saturates at relatively low field strengths.  
The \ion{Ni}{1} 676.8 nm line used by MDI and GONG is considerably less 
magnetically sensitive ($g = 1.43$), while the lines adopted by KPVT 
(\ion{Fe}{1} 868.8 nm; $g = 1.67$), SOLIS and STOP 
(\ion{Fe}{1} 630.15--630.25 nm; $g = 1.67$/2.5), and HMI 
(\ion{Fe}{1} 617.3 nm; $g = 2.5$) have intermediate sensitivity. 

Figure 1 compares the observed radial IMF strength during 1968--2021 
with the values predicted by applying a PFSS extrapolation to the 
photospheric field maps from the eight different observatories/instruments.  
For the near-Earth IMF measurements, we extracted daily values of $B_x$ 
from the OMNIWeb site\footnote{\url{http://omniweb.gsfc.nasa.gov}} 
and averaged them (without the sign) over each CR.  To calculate 
the total open fluxes, the source surface radius was fixed at 
$R_{\rm ss} = 2.5$ $R_\odot$ and $\vert B_r\vert$ was integrated over 
the source surface for each CR (Equation (1)); the results were then 
divided by $4\pi r_{\rm E}^2$ to convert them into field strengths at 1 au 
(Equation (2)).  The plotted curves were smoothed by taking 3-CR 
running averages. 

The values of $B_{\rm E}\propto\Phi_{\rm open}$ predicted by the 
observatories are spread over a wide range, sometimes differing by 
factors of up to $\sim$2--3, and they tend to be smaller than the 
measured IMF values by factors of $\sim$2--5.  As an exception from the 
general trend, however, the KPVT/SPM open flux approximately matches 
the radial IMF measurements during 1992--1997, as found earlier by 
Arge et al. (2002).  The SOLIS open flux also approaches the observed 
IMF levels during 2005--2009.  The MWO and WSO open fluxes are in 
remarkably good agreement with each other, but both are a factor of 
$\sim$2 too low compared with the observed IMF near solar minimum 
and a factor of $\sim$4--5 too low near solar maximum.\footnote{Since 
the dipole component dominates the coronal field beyond $r\sim 2$ $R_\odot$ 
(see, e.g., Hoekema 1984; Wang \& Sheeley 1988), replacing the 
chosen source surface radius of 2.5 $R_\odot$ by some other value 
$R_{\rm ss}\gtrsim 2$ $R_\odot$ would shift the open flux curves 
in Figure 1 upward or downward by a factor of 
$\sim$$(2.5\phantom{.}R_\odot/R_{\rm ss})$.} 

Because the multipole components $l$ of the photospheric field fall off 
as $r^{-(l+2)}$, the main contribution to the source surface field 
and thus to $\Phi_{\rm open}$ comes from the dipole ($l = 1$) component, 
except around the time of polar field reversal, when the quadrupole ($l = 2$) 
dominates.  The Sun's total dipole strength, $D_{\rm tot}$, is given by 
\begin{equation}
D_{\rm tot}(t) =  (D_{\rm ax}^2 + D_{\rm eq}^2)^{1/2},
\end{equation}
\begin{equation}
D_{\rm ax}(t) = \frac{3}{4\pi}\int B_r(R_\odot,L,\phi,t)\sin{L}d\Omega,
\end{equation}
\begin{equation}
D_{\rm eq}(t) = (H_1^2 + H_2^2)^{1/2},
\end{equation}
\begin{equation}
H_1(t) = \frac{3}{4\pi}\int B_r(R_\odot,L,\phi,t)\cos{L}\cos\phi d\Omega,
\end{equation}
\begin{equation}
H_2(t) = \frac{3}{4\pi}\int B_r(R_\odot,L,\phi,t)\cos{L}\sin\phi d\Omega.
\end{equation}
Figure 2 compares the values of $D_{\rm tot}$ derived for the different 
observatories, while Figure 3 shows separately the equatorial and 
axial components of the dipole vector.  As anticipated, the variation of 
the total dipole strength resembles that of the total open flux for each 
observatory (compare Figures 1 and 2).  The main difference is that 
the amplitude of the solar cycle variation is greater for $D_{\rm tot}$, 
which falls to very low values when $D_{\rm ax}$ reverses sign (see 
Figure 3(b)); at this time, $\Phi_{\rm open}$ is dominated by the 
quadrupole component of the photospheric field (CMEs also act to boost the 
observed IMF strength).  From an analysis of white-light coronagraph data 
during 2012, Wang et al. (2014) deduced that the heliospheric current sheet 
split into two conical structures lying near the equator and separated 
by $\sim$180$^\circ$ in longitude, consistent with the temporary prevalence 
of the $(l = 2,\vert m\vert = 2)$ harmonic component when $D_{\rm ax}$ 
goes through zero.

Comparing Figure 3(a) with Figure 1, we see that the large peaks in the 
IMF strength observed in 1982, 1991, 2002--2003, and 2014--2015 correspond 
to peaks in $D_{\rm eq}$.  At these times, $\vert D_{\rm ax}\vert$ 
is also rapidly increasing and approaching its final solar-minimum level 
(Figure 3(b)); as a result, the post-maximum peaks in $D_{\rm tot}$, 
$\Phi_{\rm open}$, and the observed IMF strength are characterized 
by steep rises and more gradual falloffs.  From Figures 1 and 2, 
we note that these peaks are much less prominent in the MWO and WSO 
extrapolations than in those using photospheric measurements from the 
other six observatories/instruments; this is especially clear during 
2014--2015, when the WSO total open flux and dipole strength rise to 
a plateau but the SOLIS, GONG, HMI, and STOP extrapolations all predict 
a sharp peak, consistent with the IMF observations.  The most likely 
reason for this difference is that the uncorrected MWO and WSO measurements 
overweight the fields toward the limb (or at higher latitudes), 
and thus $D_{\rm ax}$ relative to $D_{\rm eq}$ (see Section 4).

Figure 4 shows, as a function of time for each of the observatories, 
the total unsigned photospheric flux, expressed as a surface-averaged 
field strength:
\begin{equation}
B_{\rm tot}(t) = \frac{1}{4\pi}\int\vert B_r(R_\odot,L,\phi,t)\vert d\Omega.
\end{equation}
In general, $B_{\rm tot}$ is dominated by high-order multipoles of the 
photospheric field, which fall off rapidly with height and do not 
contribute to $\Phi_{\rm open}$.  Thus, agreement between the total fluxes 
measured by different observatories does not necessarily imply 
that their total dipole strengths or open fluxes agree, as may be seen 
by comparing Figure 4 with Figures 1 and 2.  To better illustrate 
this point, Figure 5(a) shows the values of $B_{\rm tot}$ derived 
for HMI plotted against those derived for GONG, while Figure 5(b) 
shows the same for $D_{\rm tot}$.  Here, each cross represents a CR.  
The HMI and GONG total photospheric fluxes display a tight linear relationship 
(with a slope of 1.04 and a correlation coefficient $cc = 0.99$), 
but the relationship between their total dipole strengths is weak and noisy 
(with a slope of 0.14 and $cc = 0.19$).  Similarly, Figure 6 shows 
that the SOLIS total fluxes are on average a factor of 1.3 larger 
than the HMI total fluxes, but their dipole strengths are as much as 
a factor of 4.1 larger.

It is apparent that, when comparing magnetograph measurements in the 
context of the global solar field, pixel-by-pixel regression or 
histogram analyses of the photospheric flux itself, as in the study of 
Riley et al. (2014), do not properly capture the differences between 
the data sets.  Instead, the focus should be on the differences between 
the lowest-order harmonic components, as in Virtanen \& Mursula (2017) 
and in the present study.

\section{The Correction Factor for the MWO and WSO Fields}
For their long-term synoptic measurements, the MWO and WSO longitudinal 
magnetographs both employ the absorption line \ion{Fe}{1} 525.0 nm, 
which has $g = 3.0$.  Since the line is also narrow, the Zeeman shift 
becomes comparable to the line width for even relatively weak fields.  
The shift is given by 
\begin{equation}
\Delta\lambda_{\rm Z} = 4.67\times 10^{-9}g\lambda^2B_{\rm los}
\phantom{.}{\rm pm}, 
\end{equation}
where $\lambda$ is in nanometers and $B_{\rm los}$ in gauss.  
The magnetograph signal may be represented by the Stokes parameter 
$V = (I_+ - I_-)/2$, where $I_+$ and $I_-$ denote the right- and 
left-circularly polarized intensities.  For unsaturated fields, 
\begin{equation}
V\simeq\Delta\lambda_{\rm Z}\frac{dI}{d\lambda}\propto 
gB_{\rm los}\frac{dI}{d\lambda} 
\end{equation}
(see, e.g., Stenflo 2013).  In the case of \ion{Fe}{1} 525.0 nm, however, 
the shifted line profile is no longer linear around the position of 
the exit slit, and Equation (10) breaks down. 

Several different saturation corrections have been proposed, most of 
which involve comparing measurements made in \ion{Fe}{1} 525.0 nm 
and \ion{Fe}{1} 523.3 nm, which, because it is three times wider than 
the 525.0 nm line and has $g = 1.3$, is assumed to remain unsaturated 
and to yield the true flux value.  In addition, the 523.3 nm line 
is less prone to weakening due to the increased temperatures associated 
with the photospheric magnetic network (Chapman \& Sheeley 1968; 
Harvey \& Livingston 1969; Hirzberger \& Wiehr 2005).

Using the MWO 150-foot tower telescope's Babcock magnetograph and 
employing a 17$^{\prime\prime}\times 17^{\prime\prime}$ scanning aperture, 
Howard \& Stenflo (1972) made alternating measurements in the two lines 
and derived a scaling factor of 
\begin{equation}
\delta^{-1} = \frac{B_{\rm los}(523.3)}{B_{\rm los}(525.0)} = 
0.48 + 1.33\cos\rho\qquad \textrm{(Howard \& Stenflo 1972)}, 
\end{equation}
where $\rho$ is the center-to-limb angle (restricted to $\rho < 60^\circ$).  
The magnetograph signal was interpreted as coming from a mixture 
of a ``filamentary'' component consisting of very narrow flux tubes 
with similar kilogauss field strengths and an ``interfilamentary'' 
component of less than 3 G; the measured flux densities reflect 
the varying filling factor of the filamentary component. 

Frazier \& Stenflo (1972) used the Kitt Peak multi-channel magnetograph 
to perform simultaneous observations in the two lines, with an 
aperture size of $2{\farcs}4\times 2{\farcs}4$.  They obtained 
\begin{equation}
\delta^{-1} = 0.845 + 1.031\cos\rho\qquad\textrm{(Frazier \& Stenflo 1972)}, 
\end{equation}
again giving a correction factor of 1.8--1.9 near disk center. 

According to Svalgaard et al. (1978), the Stanford Solar Observatory 
(now WSO) used for its \ion{Fe}{1} 525.0 nm measurements 
the same exit-slit arrangement as Howard \& Stenflo (1972) at MWO.  
The most important difference between the instruments appears to be 
the 3$^\prime$ aperture size of the WSO magnetograph, which exceeds 
those employed at MWO by a factor of 10 or more.  Svalgaard et al. 
made no comparison measurements in \ion{Fe}{1} 523.3 nm or other lines, 
but instead assumed that the average field strength in a magnetic element 
is 1500 G and argued that the corresponding reading from their magnetograph 
would be 830 G; they thus obtained a saturation correction of 1.8 
at disk center, in agreement with the result of Howard \& Stenflo (1972) 
and Frazier \& Stenflo (1972).  In addition, by tracking magnetic flux 
as it rotated across the disk, they found that $B_{\rm los}(525.0)$ varied 
as $\cos\rho$; from this, they deduced that the saturation correction 
was independent of $\rho$ and that the center-to-limb variation of 
$B_{\rm los}(525.0)$ was that expected for the simple projection of a 
radially oriented photospheric field.  They therefore concluded that 
\begin{equation}
\delta^{-1} = 1.8\qquad\textrm{(Svalgaard et al. 1978)}.
\end{equation}
We remark here that, even though MWO magnetograms recorded in the 
525.0 nm line do not show $B_{\rm los}$ falling off as $\cos\rho$ 
toward the limb, but instead remaining relatively strong, this does not 
necessarily require the photospheric field to be nonradial.

Employing the dual exit stage system of the post-1982 MWO magnetograph, 
Ulrich (1992) made simultaneous measurements in \ion{Fe}{1} 525.0 nm and 
523.3 nm with aperture sizes of $5^{\prime\prime}\times 5^{\prime\prime}$, 
$12^{\prime\prime}\times 12^{\prime\prime}$, and 
$20^{\prime\prime}\times 20^{\prime\prime}$.  Rather surprisingly, 
the saturation correction was found to be more than twice as large as the 
values derived in the earlier studies: for the two larger apertures, 
\begin{equation}
\delta^{-1} = 4.5 - 2.5\sin^2\rho\qquad\textrm{(Ulrich 1992)}, 
\end{equation}
with the $5^{\prime\prime}\times 5^{\prime\prime}$ aperture giving 
somewhat smaller values.  As discussed below, Ulrich et al. (2009) 
later argued that this scaling factor should be reduced to 
\begin{equation}
\delta^{-1} = 4.15 - 2.82\sin^2\rho\qquad\textrm{(Ulrich et al. 2009)}.
\end{equation}

Using the STOP magnetograph at the Syan Solar Observatory, 
Demidov \& Balthasar (2009) obtained the full Stokes $I$ and $V$ profiles 
for both the 525.0 nm and 523.3 nm lines.  They showed that the separation 
between the peaks of their Stokes $V$ profiles (normalized to the 
continuum intensity $I_{\rm c}$) is generally not a measure of the 
field strength.  They also found that the Stokes $V/I_{\rm c}$ 
peak positions were close to the steepest parts of the 
corresponding Stokes $I$ profiles for both lines.  To derive 
$B_{\rm los}(523.3)/B_{\rm los}(525.0)$, they took the ratio of 
the $V/I_{\rm c}$ peaks for the two lines, after averaging between 
the amplitudes of the blue- and red-wing peaks for each line 
and dividing out the respective Land{\'e} factors (see Equation (10)).  
The result for 10$^{\prime\prime}$ spatial resolution was 
\begin{equation}
\delta^{-1} = 1.74 - 2.43\cos\rho + 3.43\cos^2\rho\qquad\textrm{(Demidov \& Balthasar 2009)}, 
\end{equation}
with the disk center value of 2.74 increasing to 2.83 when the 
resolution was decreased to 100$^{\prime\prime}$. 

As may be seen from Figures 1 through 4, the values of $\Phi_{\rm open}$, 
$D_{\rm tot}$, $D_{\rm eq}$, $D_{\rm ax}$, and $B_{\rm tot}$ derived 
from the MWO and WSO photospheric field maps are generally in 
excellent agreement with each other.  This suggests that the MWO and WSO 
measurements require a similar correction factor (contrary to the 
conclusion of Riley et al. 2014), and that the constant scaling factor 
of 1.8 found by Svalgaard et al. (1978) using the WSO magnetograph 
cannot be correct if any of the $\rho$-dependent scaling factors 
derived at the other observatories are correct.  One possible source 
of the disagreement is the much larger scanning aperture employed 
at WSO: it may be that the proportion of weak ``interfilamentary'' fields 
within the aperture increases systematically toward the limb, 
so that the $\cos\rho$ dependence of the measured line-of-sight field 
is partly coincidental.

Ulrich et al. (2009) presented a detailed analysis of the 
saturation corrections derived using the \ion{Fe}{1} 523.3 nm line.  
Their main result was that the measured field depends sensitively on 
where in the 523.3 nm line wings the exit slits are placed.  
To demonstrate this, they used the MWO magnetograph (with a 
$12^{\prime\prime}\times 12^{\prime\prime}$ aperture) to scan 
the 523.3 nm line profile in both polarization states at a 
sampling interval of 0.53 pm, focusing their analysis on a sequence 
of profiles that they obtained for a plage region on 2007 July 13.  
Their Figure 6 shows how the Zeeman shift may be derived from the 
difference ($2\delta\lambda_{\rm bi}$) between the line bisector 
positions ($\lambda_{\rm bi}^+$, $\lambda_{\rm bi}^-$), defined for 
each state of polarization as being midway between a given pair of 
equal-intensity points on the red and blue wings of the line.  
The field strength is then obtained by setting 
$\Delta\lambda_{\rm Z} = \delta\lambda_{\rm bi}$ in Equation (9) above.  
However, the result depends sensitively on where along the 
523.3 nm line profile the bisector is calculated, because the red 
and blue wings are asymmetric and their shapes also differ between 
the two polarization states.

The dependence of $B_{\rm los}(523.3)$ on $\Delta\lambda$, the position 
on the 523.3 nm line wing used to determine the bisector locations, 
is shown in Figure 7, where the curve is the same as that plotted 
in Figure 7 of Ulrich et al. (2009).  We have superposed on this plot 
the center positions of the exit slits employed by Howard \& Stenflo (1972), 
Frazier \& Stenflo (1972), Ulrich (1992), Ulrich et al. (2009), and 
Demidov \& Balthasar (2009).  For the plage area under observation, 
$B_{\rm los}(523.3)$ ranges from a maximum value of $\sim$450 G 
at $\Delta\lambda\sim 8$ pm to a minimum value of $\sim$160 G at 
$\Delta\lambda\gtrsim 25$ pm.  This sensitivity to where the 523.3 nm 
line profile is sampled may be one of the main reasons for the 
large differences among the 525.0 nm saturation corrections.  
In particular, Ulrich (1992) centered the 523.3 nm exit slit at 
$\Delta\lambda = 8.7$ pm, near the peak of the curve in Figure 7, 
whereas Howard \& Stenflo (1972) chose $\Delta\lambda = 14$ pm; 
for the case shown in the figure, the derived values of $B_{\rm los}(523.3)$ 
would be $\sim$444 G and $\sim$275 G, respectively, representing a factor 
of $\sim$1.61 difference.  Similarly, Frazier \& Stenflo (1972) centered 
their 523.3 nm slit at $\Delta\lambda = 16.2$ nm, which corresponds to 
$B_{\rm los}(523.3)\sim 227$ G according to Figure 7, a factor of 
$\sim$1.95 smaller than would be obtained with the slit positioned 
at 8.7 nm.  Demidov \& Balthasar (2009) used the bisector method 
to derive an estimate of the 525.0 nm saturation correction near 
disk center; in their illustrative case, they centered their ``virtual'' 
523.3 nm slit at $\Delta\lambda = 15.53$ pm and obtained 
$B_{\rm los}(523.3)/B_{\rm los}(525.0)\sim 2.46$. 

Figure 7 raises the question of where exactly the 523.3 nm line profile 
should be sampled.  Ulrich et al. (2009) suggested that the exit slit 
should be placed as close to the line center as possible, so that 
the field is measured at greater heights where it becomes more uniform.  
They chose $\Delta\lambda = 2.9$ pm; because of the dip in the 
$B_{\rm los}(523.3)$ curve at small values of $\Delta\lambda$, 
the resulting saturation correction is then somewhat reduced relative to 
that derived by Ulrich (1992), who chose $\Delta\lambda = 8.7$ pm 
(compare Equations (14) and (15)).  However, the earlier (fortuitous) choice 
close to the peak of the curve may have been more appropriate, for the 
following reasons.  Table 2 of Ulrich et al. (2009) gives the 
heights of formation of the 525.0 nm and 523.3 nm lines as a function 
of wavelength position and $\rho$, calculated using the Harvard--Smithsonian 
Reference Atmosphere and the radiative transfer methods described in 
Caccin et al. (1977), under the assumption of LTE.  The radiation at the 
525.0 nm slit position adopted by Ulrich (1992), $\Delta\lambda = 3.9$ pm, 
comes from a height of $h = 185$ km at disk center.  For the 523.3 nm line 
(again for $\rho = 0$), the estimated heights of formation corresponding to 
$\Delta\lambda = 0.9$, 2.8, 8.4, 10.2, and 17.7 pm are $h = 550$, 482, 
145, 122, and 92 km, respectively.  Thus, sampling the 523.3 nm line 
at $\Delta\lambda = 2.9$ pm means that the signal originates from 
much greater heights than that recorded in the 525.0 nm line.  
Conversely, placing the slit at $\Delta\lambda = 14$ pm (following 
Howard \& Stenflo 1972) or 16.2 pm (following Frazier \& Stenflo 1972) 
would cause the 523.3 nm signal to originate at substantially lower heights 
than the 525.0 nm signal.  The original choice $\Delta\lambda = 8.7$ pm 
makes the heights of formation of the 523.3 and 525.0 nm signals 
comparable to each other (see also Figure 4 in Ulrich et al. 2002).

A second reason for not placing the exit slit near the line center 
is that the photospheric field fans out with height on a horizontal scale 
comparable to a supergranule radius, or $\sim$20$^{\prime\prime}$ (the 
canopy effect).  Since the scanning apertures used with the MWO magnetograph 
have dimensions $\lesssim 20^{\prime\prime}\times 20^{\prime\prime}$, 
some of the flux may escape sideways out of the aperture 
if the signal originates from heights of order 500 km, as is the 
case if $\Delta\lambda\lesssim 3$ pm.  This might also be a reason 
for the decrease in $B_{\rm los}(523.3)$ near line center.

It is unclear to us why $B_{\rm los}(523.3)$ decreases steeply 
beyond $\Delta\lambda\sim 10$ pm, corresponding to heights below 
$\sim$120 km at disk center.  Plowman \& Berger (2020a,b,c) have suggested 
that unresolved granular structure at the photosphere causes 
magnetographs to underweight the strong flux in the dark intergranular lanes 
and to overweight the weak flux in the bright granule centers.  However, 
G. J. D. Petrie (2021, private communication), using a different 
radiation MHD model for a sunspot and its surroundings, found almost no 
systematic correlation between intensity and field strength.

The shape of the $B_{\rm los}(523.3)$ versus $\Delta\lambda$ curve 
in Figure 7 has recently been confirmed by one of us (JWH) using 
Stokes $I$ and $V$ polarimetry data obtained with the McMath--Pierce/NSO 
Fourier Transform Spectrometer (FTS), as part of an ongoing survey of 
Zeeman splitting of a variety of spectral lines (J. W. Harvey, 
in preparation).\footnote{The FTS is described by Brault (1978) and 
some specific observations have been presented by Stenflo et al. (1984).}  
On 1979 April 30, a spotless, unipolar region was observed at 
$\cos\rho = 0.91$ using a 10$^{\prime\prime}$ circular aperture 
for a total integration time of 35 minutes.  The wavelength range covered 
was $\sim$100 nm wide and centered at 505 nm, and included more than 
2500 simultaneously recorded spectral lines.  Line bisectors are here 
calculated for the 523.3 and 525.0 nm lines from their $I\pm V$ spectra.  
One-half the wavelength difference between the bisectors is scaled to 
$B_{\rm los}$ by Equation (9) and plotted in Figure 8 as a function of 
one-half the $I(\lambda)$ line width.  Except for the different field strengths 
associated with the different plage regions under observation, the shapes 
of the 523.3 nm curves in Figures 7 and 8 are strikingly similar, 
despite the completely different instruments and observing techniques 
used to obtain them (to make this even clearer, the Figure 7 curve 
has been scaled down and replotted in red in Figure 8).  Note that 
the $B_{\rm los}$ curves have opposite slopes for the two lines, 
increasing by a factor of more than two moving outward along the 
525.0 nm wings, but decreasing by a factor of three in the wings of 523.3 nm 
beyond the peak at $\Delta\lambda\sim 8$ pm.  This at least partly explains  
why the values of $\delta^{-1}$ found by different investigators are 
so discordant.  Because Howard \& Stenflo (1972), Frazier \& Stenflo (1972), 
and Demidov \& Balthasar (2009) placed their 525.0 nm exit slits somewhat 
farther away from line center than Ulrich (1992) (i.e., at 4.65, 5.5, and 
6.21 pm rather than at 3.9 pm), the net result may have been to 
decrease further the $B_{\rm los}(523.3)/B_{\rm los}(525.0)$ ratios 
that they derived compared with that obtained by Ulrich (1992).

In addition to the line profile measurements that were the basis 
for their (and our) Figure 7, Ulrich et al. (2009) described a new set of 
MWO observations during 2007 April--May, in which they obtained pairs 
of magnetograms with the 525.0 nm slit position fixed at $\pm$3.9 pm 
(as in Ulrich 1992) and the 523.3 nm line sampled at $\pm$0.9, $\pm$2.9, 
$\pm$8.4, $\pm$10.2, and $\pm$17.7 pm.  The scatter diagrams in their 
Figure 3 show the relationship between $B_{\rm los}(523.3)$ and 
$B_{\rm los}(525.0)$ for different $\rho$, for the case where the 
523.3 nm slit is centered at $\pm$8.4 nm (near the peak of the curve 
in Figure 7).  Near disk center ($0.0 < \rho < 0.4$), the regression line 
has a slope of $5.525\pm 0.107$, while near the limb ($0.8 < \rho < 0.95$) 
the slope is $2.757\pm 0.083$.  The center-to-limb variation of 
$B_{\rm los}(523.3)/B_{\rm los}(525.0)$ may then be approximated as 
\begin{equation}
\delta^{-1} = 5.5 - 2.8\sin^2\rho. 
\end{equation}
This correction factor is even larger than that found by Ulrich (1992) 
employing an earlier MWO spectrograph system, and substantially exceeds 
the result obtained when the 523.3 nm slit is placed at $\pm$2.9 pm 
(Equation(15)).

\section{Comparison with the Observed Radial IMF Variation}
We now apply the three \ion{Fe}{1} 525.0 nm saturation corrections from 
Ulrich (1992) and Ulrich et al. (2009), given by Equations (14), (15), and 
(17), to the MWO and WSO photospheric field maps, and recalculate the 
total open fluxes using the PFSS model with $R_{\rm ss} = 2.5$ $R_\odot$.  
The results are plotted in Figure 9, along with the near-Earth radial IMF 
variation during 1968--2021; all curves have been smoothed by taking 3-CR 
running means.  The open fluxes obtained using the different corrections 
are all reasonably well correlated with the IMF variation, with $cc$ 
ranging from 0.71 to 0.76 when the MWO fields are extrapolated and 
from 0.79 to 0.86 when the WSO maps are employed.  However, the values 
of $B_{\rm E}$ predicted with the $\delta^{-1} = (5.5 - 2.8\sin^2L)$ 
scaling factor (where we have replaced the center-to-limb angle $\rho$ 
by latitude $L$) are systematically too high before 1998, whereas the 
$\delta^{-1} = (4.15 - 2.82\sin^2L)$ scaling predicts values that 
are systematically too low (by an average of $\sim$29\%) throughout 
the 53 yr interval.  The best overall match to the observed IMF 
is obtained using the original $\delta^{-1} = (4.5 - 2.5\sin^2L)$ correction 
derived by Ulrich (1992).  However, this scaling (like that for which 
$\delta^{-1} = 4.15$ at $L = 0$) gives values of $B_{\rm E}$ 
that are too low during the rising and maximum phases of the 
last two sunspot cycles.

It should be noted that all three correction factors give rise to 
prominent peaks in $\Phi_{\rm open}$  during 1982, 1991, 2002--2003, and 
2014--2015, in agreement with the observed IMF variation.  In contrast, 
the uncorrected values of the MWO and WSO open fluxes (Figure 1) 
and total dipole strengths (Figure 2) did not reproduce these peaks; 
they now appear because $\delta^{-1}(L)$ weights the low-latitude fields 
twice as much as the polar fields, increasing the strength of 
$D_{\rm eq}$ relative to $D_{\rm ax}$.  

An inspection of Figure 9(b) also shows that, whichever saturation 
correction is adopted, the open fluxes derived from the WSO maps 
are systematically smaller relative to the observed IMF after $\sim$1998, 
as compared with earlier years (cf. Virtanen \& Mursula 2017).  
This suggests that there may have been a decrease in the sensitivity 
of the WSO magnetograph sometime around 1998.

Wang \& Sheeley (2015) estimated the contribution of ICMEs to the 
radial IMF strength at Earth during 1996--2015, including 457 events 
listed in the Richardson--Cane catalog\footnote{\url{
http://www.srl.caltech.edu/ACE/ASC/DATA/level3/icmetable2.htm}} 
and assigning to each ICME a radial field strength extracted from the 
OMNIWeb database.  They found that, averaged over the interval 
1999--2002 (2011--2014), ICMEs accounted for $\sim$23\% ($\sim$18\%) 
of the observed $B_{\rm E}$.  Figure 10 shows the effect of adding 
the contribution of ICMEs to the MWO and WSO open fluxes, for the 
case where the saturation correction has the form 
$\delta^{-1} = (4.5 - 2.5\sin^2L)$.  The observed IMF variation is 
sufficiently well reproduced as to suggest that this scaling factor 
and the inclusion of ICMEs provide a reasonable solution to the 
open flux problem.
   
\section{Summary and Discussion}
In addressing the open flux problem, we have focused on what we consider 
to be by far the ``weakest link'': the magnetograph measurements 
themselves.  Although Linker et al. (2017) and Riley et al. (2019) 
have argued that it is implausible that all of the different observatories 
should be systematically underestimating the photospheric flux, 
our analysis of the MWO and WSO measurements suggests that the uncertainties 
involved in interpreting the magnetograph signals are greater than 
sometimes supposed.  We now summarize our conclusions. 

1. When PFSS extrapolations with source surface at 2.5 $R_\odot$ 
are applied to (unmodified) photospheric field maps from MWO, WSO, KPVT, 
MDI, SOLIS, GONG, HMI, and STOP, the total open fluxes underestimate 
the observed radial IMF strength by factors of $\sim$2--5.  An exception 
is the KPVT/SPM open flux, which briefly matched the IMF level 
during 1992--1997 (cf. Arge et al. 2002); the SOLIS values were also 
reasonably close to the observed level during 2005--2009.

2. Locating the source surface well inside $r\sim 1.5$ $R_\odot$ 
(see, e.g., Badman et al. 2020) would reduce the discrepancies, but 
would result in open field areas much larger than observed coronal holes 
and in closed field regions that do not extend outward as far as 
the LASCO C2 helmet streamers, whose cusps are located at and beyond 
$r\sim 2.5$ $R_\odot$.

3. The total open flux is determined by the lowest-order multipoles 
of the photospheric field, in particular the dipole and (during polar 
field reversal) quadrupole components.  Because the total and local 
photospheric fluxes are dominated by high-order multipoles, agreement 
between the values of $B_{\rm tot}$ or $B_{\rm los}(R_\odot,L,\phi)$ 
measured by different observatories does not imply that their 
dipole strengths or open fluxes agree.  For example, the scatter plots 
in Figure 5 show that the HMI and GONG values of $B_{\rm tot}$ have 
a correlation of 0.99, with the regression line having a slope of 1.04, 
whereas their total dipole strengths $D_{\rm tot}$ have a correlation of 0.19 
and a regression line slope of only 0.13.  Similarly, the SOLIS values 
of $B_{\rm tot}$ are $\sim$1.3 times higher than the HMI values, but their 
total dipole strengths are as much as $\sim$4.1 times higher (Figure 6).

4. The values of $\Phi_{\rm open}$, $D_{\rm tot}$, and $B_{\rm tot}$ 
derived from the uncorrected MWO and WSO photospheric field maps 
are in remarkably good agreement with each other, indicating that 
they require approximately the same correction factor.  Both observatories 
predict much weaker post-maximum peaks in $D_{\rm tot}$ and 
$\Phi_{\rm open}$ than the other observatories, suggesting that 
they underestimate the contribution of the equatorial dipole component,  
the main source of these peaks (which are also present in the observed IMF).  
A correction factor that is larger at low latitudes than near the poles 
would boost the relative strength of $D_{\rm eq}$ and make the MWO 
and WSO peaks more prominent.

5. The long-term synoptic measurements at MWO and WSO both employ the 
\ion{Fe}{1} 525.0 nm absorption line, which has Land{\'e} factor 
$g = 3.0$.   Most of the correction factors that have been proposed 
are based on comparisons with measurements in the non-saturating 
\ion{Fe}{1} 523.3 nm line ($g = 1.3$).  However, the values of 
$\delta^{-1} = B_{\rm los}(523.3)/B_{\rm los}(525.0)$ derived by 
Ulrich (1992) and Ulrich et al. (2009) are a factor of $\sim$2 higher 
than those found by Howard \& Stenflo (1972), Frazier \& Stenflo (1972), 
and Demidov \& Balthasar (2009). 

6. The constant factor of 1.8 correction obtained by Svalgaard et al. (1978) 
using the WSO magnetograph is the only one that is independent of 
center-to-limb angle $\rho$ and that is not based on comparisons with 
the 523.3 nm line (they simply assumed that the maximum signal of 830 G 
recorded with their instrument in the 525.0 nm line corresponded to 
an actual field strength of 1500 G).  Their finding that 
$B_{\rm los}(525.0)\propto\cos\rho$ might be an artifact of the 
unusually wide (3$^\prime$) scanning aperture of the WSO instrument, 
if the larger and larger surface areas it averages over toward the limb 
include an increasing fraction of weak ``interfilamentary'' fields.

7. As shown by Ulrich et al. (2009), the field strength obtained using 
\ion{Fe}{1} 523.3 nm depends sensitively on where along the line wing the 
exit slit is placed: $B_{\rm los}(523.3)$ peaks at $\Delta\lambda\sim 8$ pm 
but falls by a factor of $\sim$2 when $\Delta\lambda\sim 16$ pm.  
By centering the slit at $\Delta\lambda = 8.7$ pm, Ulrich (1992) 
found that $\delta^{-1} = 4.5$ at disk center, roughly twice the 
values obtained by Howard \& Stenflo (1972), Frazier \& Stenflo (1972), 
and Demidov \& Balthasar (2009), who centered their slits at 
$\Delta\lambda = 14$--16.2 pm.

8. We have argued that the 523.3 nm line profile should be sampled 
at the wavelength position whose associated height of origin is the same 
as that of the position where the 525.0 nm profile is sampled.  
Ulrich (1992) and Ulrich et al. (2009) placed their 525.0 nm exit slits 
at $\Delta\lambda = 3.9$ pm.  According to Table 2 of Ulrich et al. (2009), 
525.0 nm $\pm$ 3.9 pm corresponds to a formation height $h = 185$ km 
at disk center, reasonably close to the height of 145 km corresponding 
to 523.3 nm $\pm$ 8.4 pm.\footnote{Figure 4 in Ulrich et al. (2002) shows 
the heights of formation for 525.0 nm $\pm$ 3.9 pm and 523.3 nm $\pm$ 8.8 pm 
to be similar at all center-to-limb angles, with $h\sim 180$ km 
at $\rho = 0$ in both cases.}  The radiation at 523.3 nm $\pm$ 14--16.2 pm 
(the slit positions adopted by Howard \& Stenflo, Frazier \& Stenflo, and 
Demidov \& Balthasar) comes from lower in the atmosphere ($h\sim 100$ km).

9. Ulrich et al. (2009) suggested that the 523.3 nm profile should be 
sampled not at $\Delta\lambda\sim 8$ pm but closer to the line center, 
where the field becomes less structured and more uniform; they chose 
$\Delta\lambda = 2.9$ pm, corresponding to a height of $\sim$480 km.  
Because $B_{\rm los}(523.3)$ falls to a local minimum at 
line center (Figure 7), the resulting correction factor 
$\delta^{-1} = (4.15 - 2.82\sin^2\rho)$ is somewhat smaller than that 
derived by Ulrich (1992) by taking $\Delta\lambda = 8.7$ pm, 
$\delta^{-1} = (4.5 - 2.5\sin^2\rho)$.  However, the formation height 
corresponding to the revised slit position is no longer consistent with 
the height of 185 km corresponding to the 525.0 nm slit position 
at $\Delta\lambda = 3.9$ pm.  Moreover, the total flux within the 
scanning aperture at $h\sim 480$ km may be less than that at the 
photospheric level because of the fanning-out of the field lines; 
indeed, this could be one reason for the dip in $B_{\rm los}(523.3)$ 
around line center.  These arguments support the idea that the 
523.3 nm line should be sampled near the peak of the curve in Figure 7, 
as was done by Ulrich (1992).

10. The shape of the $B_{\rm los}(523.3)$ curve in Figure 7 has been 
confirmed by one of us (JWH) by applying the line bisector method to 
spectropolarimetric data from NSO/FTS (Figure 8).  The latter figure 
shows that $B_{\rm los}(525.0)$ is also a function of line profile position, 
but with the field strengths increasing rather than decreasing when
moving outward along the wings.

11. Figure 9 shows that the best overall fit to the observed radial IMF 
variation during 1968--2021 is obtained by applying the 
$\delta^{-1} = (4.5 - 2.5\sin^2L)$ scaling factor to either the MWO 
or the WSO photospheric field maps.  The fit is further improved by 
including ICMEs from the Richardson--Cane catalog, which contribute 
$\sim$20\% of the IMF flux during the rising and maximum phases 
of the solar cycle (see Figure 10).  We therefore suggest that the 
Ulrich (1992) saturation correction supplemented by ICMEs provides 
a plausible solution to the open flux problem.

An important question that remains to be answered is why the derived values 
of $B_{\rm los}(523.3)$ fall off steeply beyond $\Delta\lambda\sim 10$ pm, 
corresponding to heights $\lesssim$100 km.  One possibility is that the 
circular polarization in the outer line wings is weakened through collisions, 
perhaps analogous to the collisional damping effect first discussed 
by Zanstra (1941) in the context of the depolarization of the 
\ion{Ca}{1} 422.7 nm line wings near the solar limb.  Such questions 
may serve as a reminder that the interpretation of the magnetograph signals 
from the variety of spectral lines listed in Table 1 is likely to be 
far from straighforward.

We are greatly indebted to L. Bertello, J. T. Hoeksema, Y. Liu, 
G. J. D. Petrie, A. A. Pevtsov, V. M. Pillet, N. R. Sheeley, Jr., 
L. Svalgaard, and A. G. Tlatov for helpful email discussions.  The STOP maps 
were kindly made available to us by Dr. Tlatov.  We have also utilized 
data from the NSO Integrated Synoptic Program, which is operated by 
the Association of Universities for Research in Astronomy under a 
cooperative agreement with NSF.  This work was supported by NASA and 
the Office of Naval Research.

\newpage

\begin{deluxetable}{ccccc}
\rotate
\tabletypesize{\scriptsize}
\tablewidth{0pt}
\tablecaption{Photospheric Magnetic Field Maps Used in This Study}
\tablehead{
\colhead{Observatory/Instrument\tablenotemark{a}} & \colhead{Period Covered} & 
\colhead{Map Dimensions\tablenotemark{b}} & \colhead{Spectral Line} & 
\colhead{Land{\'e} $g$}}
\startdata
Mount Wilson Observatory (MWO)\tablenotemark{c}{\dotfill}& 1967--2013 (CR 1516--2132) & 91$\times$34 & Fe I 525.0 nm & 3.00 \\

Wilcox Solar Observatory (WSO){\dotfill}& 1976--present (CR 1642--present) & 72$\times$30 & Fe I 525.0 nm & 3.00 \\

NSO Kitt Peak Vacuum Telescope Spectromagnetograph (KPVT/SPM){\dotfill}& 1992--2003 (CR 1863--2007) & 360$\times$180 & Fe I 868.8 nm & 1.67 \\

NSO SOLIS Vector Spectromagnetograph (SOLIS/VSM){\dotfill}& 2003--2017 (CR 2007--2195) & 360$\times$180 & Fe I 630.15/630.25 nm & 1.67/2.50 \\

NSO Global Oscillation Network Group (GONG){\dotfill}& 2006--present (CR 2047--present) & 360$\times$180 & Ni I 676.8 nm\tablenotemark{d} & 1.43 \\

Michelson Doppler Interferometer (MDI)\tablenotemark{e}{\dotfill}& 1996--1998, 1999--2010 (CR 1909--1937, 1947--2104) & 3600$\times$1080 & Ni I 676.8 nm\tablenotemark{d} & 1.43 \\

Helioseismic and Magnetic Imager (HMI)\tablenotemark{f}{\dotfill}& 2010--present (CR 2097--present) & 720$\times$360 & Fe I 617.3 nm\tablenotemark{d} & 2.50 \\

Kislovodsk Solar Telescope for Operative Predictions (STOP)\tablenotemark{g}{\dotfill}& 2014--present (CR 2152--present) & 720$\times$360 & Fe I 630.15/630.25 nm & 1.67/2.50 

\enddata
\tablenotetext{a}{For a historical review of the different magnetographs and their synoptic datasets, see Pevtsov et al. (2021).} 
\tablenotetext{b}{Number of pixels in longitude and sine latitude (except for the STOP maps, which have equal latitude spacing).  The MWO and WSO maps were subsequently interpolated to 5$^\circ$ pixels in longitude and latitude, while the remaining maps were rebinned to 1$^\circ$ pixels in longitude and latitude.} 
\tablenotetext{c}{During 1982--1988, a defective circular polarizer was used to measure the reduction in the MWO magnetograph signal caused by a low-pass filter; based on our estimate of the actual weakening of the signal, we have multiplied the MWO fields during CR 1721--1807 by a factor of 1.42.} 
\tablenotetext{d}{For a discussion of the use of the Fe I 617.3 nm and Ni I 676.8 nm lines in magnetic and Doppler velocity measurements, see Norton et al. (2006).} 
\tablenotetext{e}{The MDI maps are Level 1.8.2, which includes a scaling factor of $\sim$1.7 based on the MWO vs MDI comparison of Tran et al. (2005).} 
\tablenotetext{f}{According to Liu et al. (2012), scaling the HMI fields upward by 1.4 makes them comparable to the Level 1.8.2 MDI fields.} 
\tablenotetext{g}{See Berezin \& Tlatov (2020), and references therein.} 
%\tablecomments{}
\end{deluxetable}
\newpage

\begin{figure}
\vspace{-5.5cm}
\centerline{\includegraphics[width=42pc]{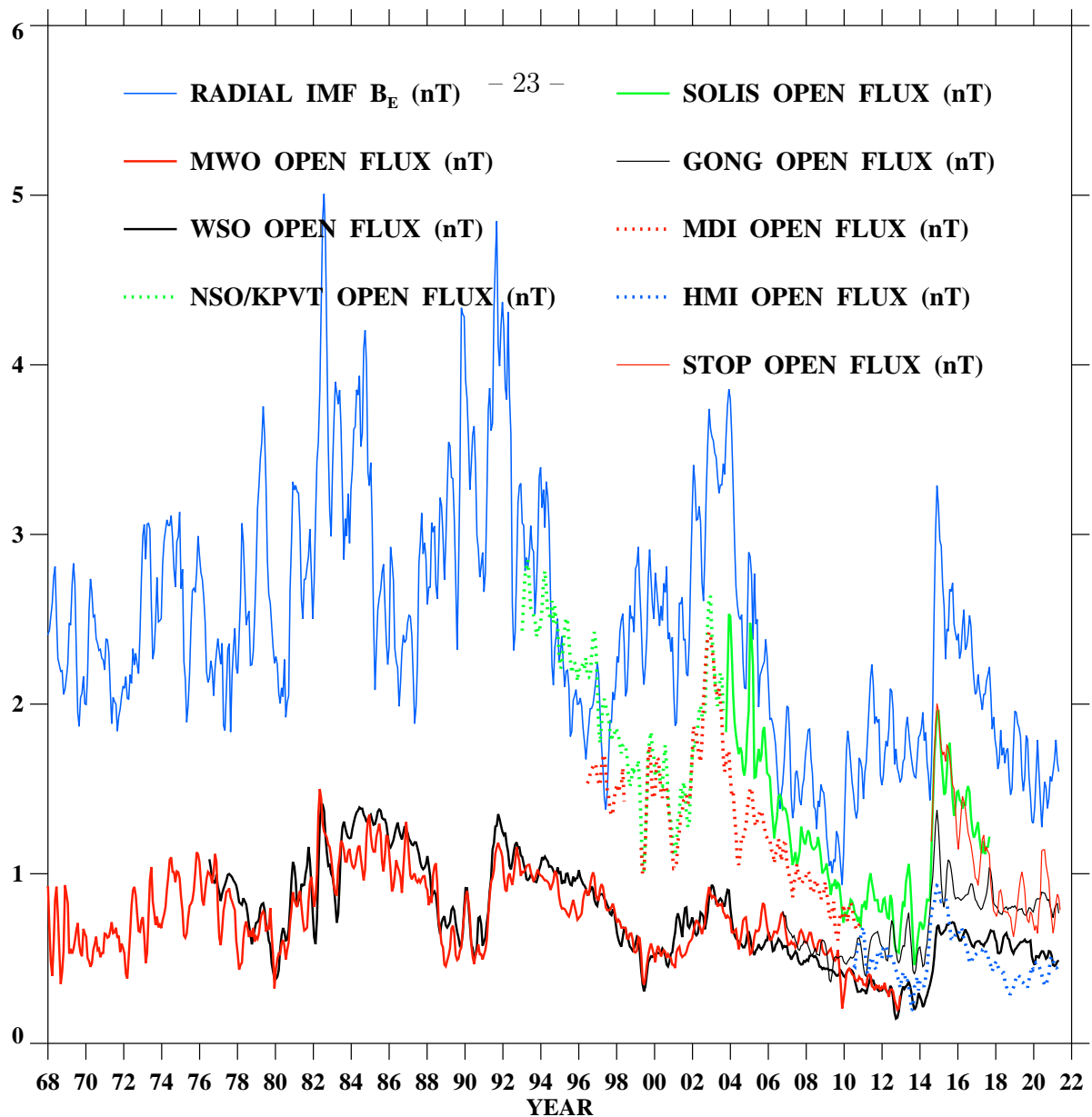}}
\vspace{-3.7cm}
\caption{Comparison between the near-Earth radial IMF strength measured 
during 1968--2021 and the total open fluxes derived by applying a 
PFSS extrapolation to photospheric field maps from MWO, WSO, KPVT/SPM, 
SOLIS, GONG, MDI, HMI, and STOP.  The source surface radius was fixed at 
$R_{\rm ss} = 2.5$ $R_\odot$, $B_r$ was matched to the photospheric field 
on the assumption that it is radially oriented, and the total unsigned 
flux crossing the source surface was converted into a field strength 
at 1 au by dividing by $4\pi r_{\rm E}^2$.  Daily values of $B_x$ 
measured near Earth were extracted from the OMNIWeb database and 
averaged without the sign over successive CRs.  The MWO and WSO maps 
were interpolated to 72 longitude pixels by 36 latitude pixels, 
while the remaining maps were regridded to dimensions of 360$\times$180; 
no other corrections (other than for line-of-sight projection) were applied 
to the maps after downloading them from the observatory websites.  Here and 
in the next three figures, all curves represent 3-CR running averages.}
\end{figure}

\clearpage
\begin{figure}
\vspace{-6.0cm}
\centerline{\includegraphics[width=42pc]{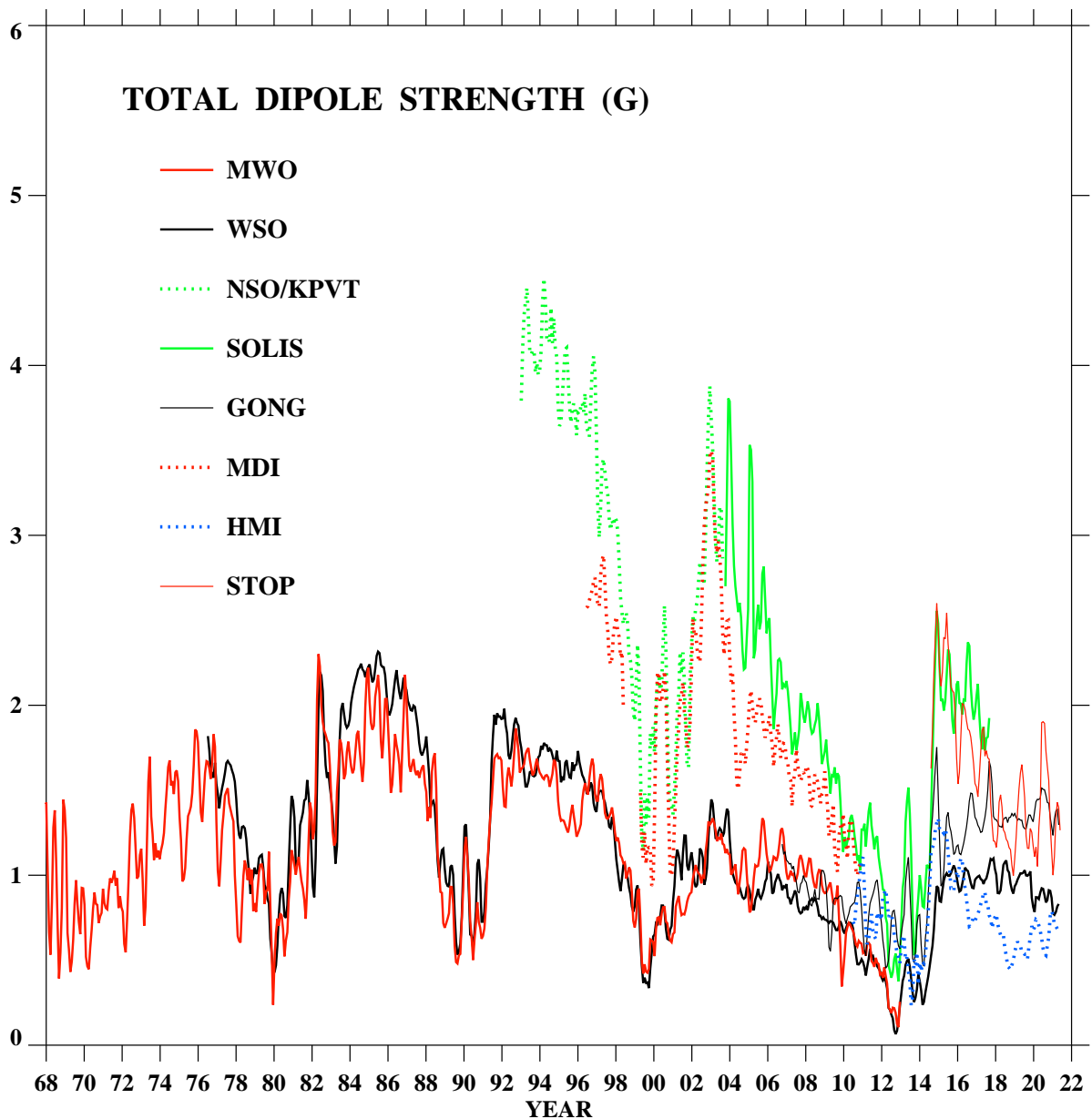}}
\vspace{-3.7cm}
\caption{Comparison of the total dipole strengths $D_{\rm tot}$ 
derived from the photospheric field maps from MWO, WSO, KPVT/SPM, 
SOLIS, GONG, MDI, HMI, and STOP.  The dipole component provides 
the main contribution to the source surface field and open flux, 
except during polar field reversal, when the quadrupole component 
dominates.  Note that MWO and WSO show much weaker peaks in $D_{\rm tot}$ 
during 2002--2003 and 2014--2015 than the other observatories; these peaks 
also appear in the observed IMF variation (see Figure 1).}
\end{figure}

\clearpage
\begin{figure}
\vspace{-3.0cm}
\centerline{\includegraphics[width=42pc]{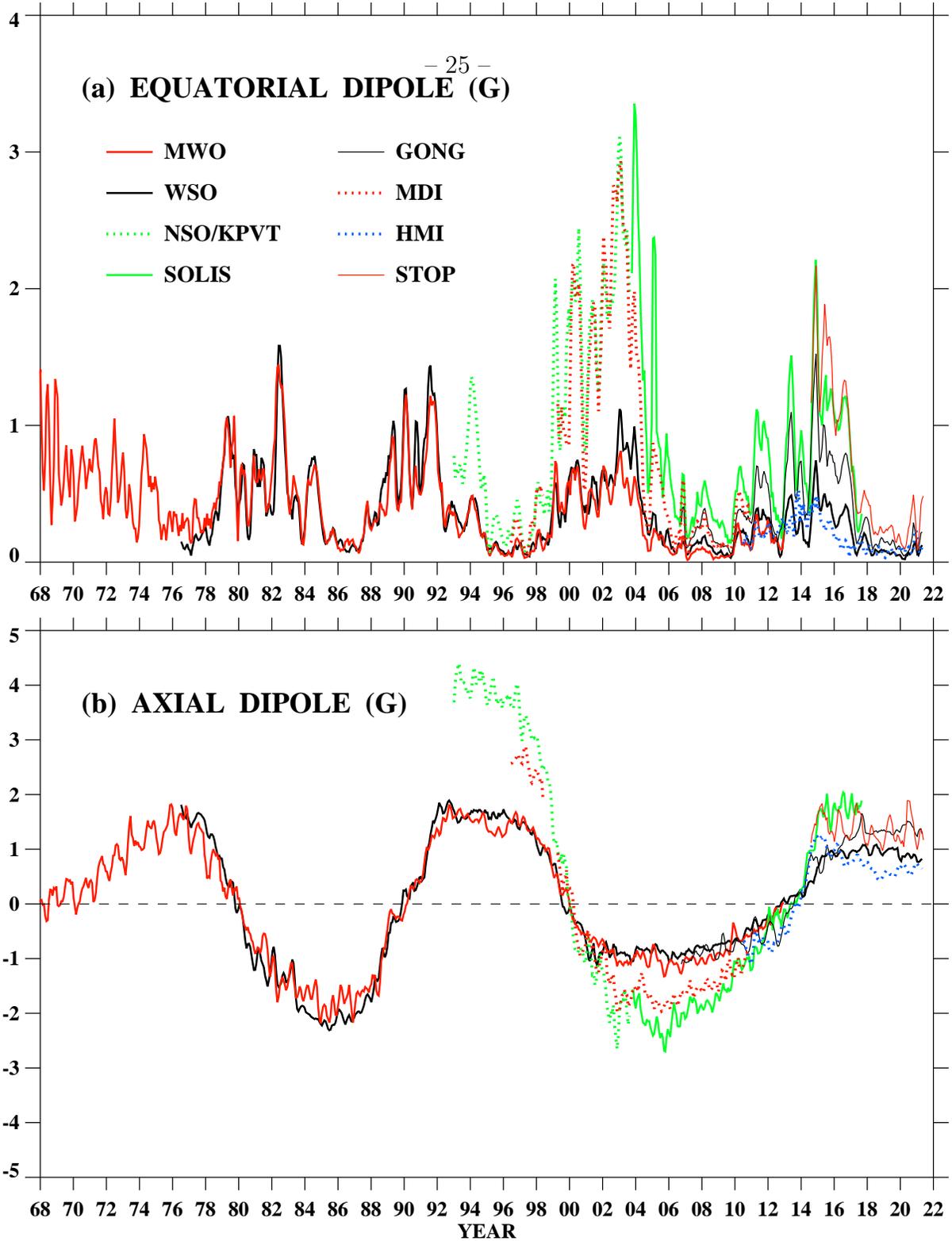}}
\vspace{-1.3cm}
\caption{(a) Variation of $D_{\rm eq}$, the equatorial dipole or 
$(l=1,\vert m\vert=1)$ component of the photospheric field.  (b) Variation 
of $D_{\rm ax}$, the axial dipole or $(l=1,m=0)$ component of the 
photospheric field.  Although the amplitude of the axial dipole variation 
shows a wide scatter between observatories, the reversal times are in 
good agreement.}
\end{figure}

\clearpage
\begin{figure}
\vspace{-6.0cm}
\centerline{\includegraphics[width=42pc]{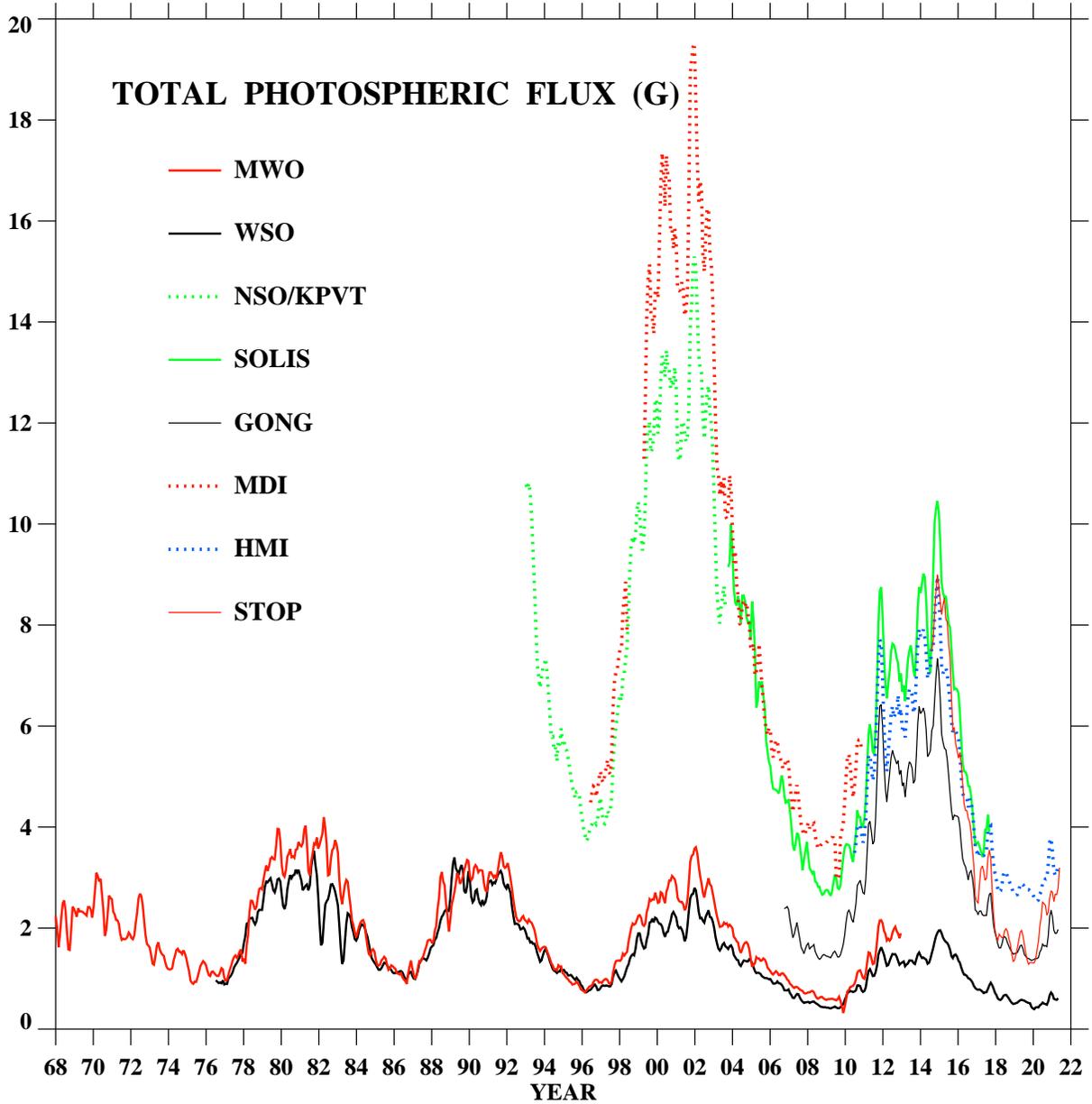}}
\vspace{-3.7cm}
\caption{Variation of $B_{\rm tot}$, the total unsigned photospheric flux 
averaged over the solar surface.  Agreement between total fluxes may 
(as in the case of MWO and WSO) or may not (as in the case of SOLIS and HMI) 
entail agreement between open fluxes or dipole strengths (compare Figure 4 
with Figures 1 to 3).}
\end{figure}

\clearpage
\begin{figure}
\vspace{-5.0cm}
\centerline{\includegraphics[width=42pc]{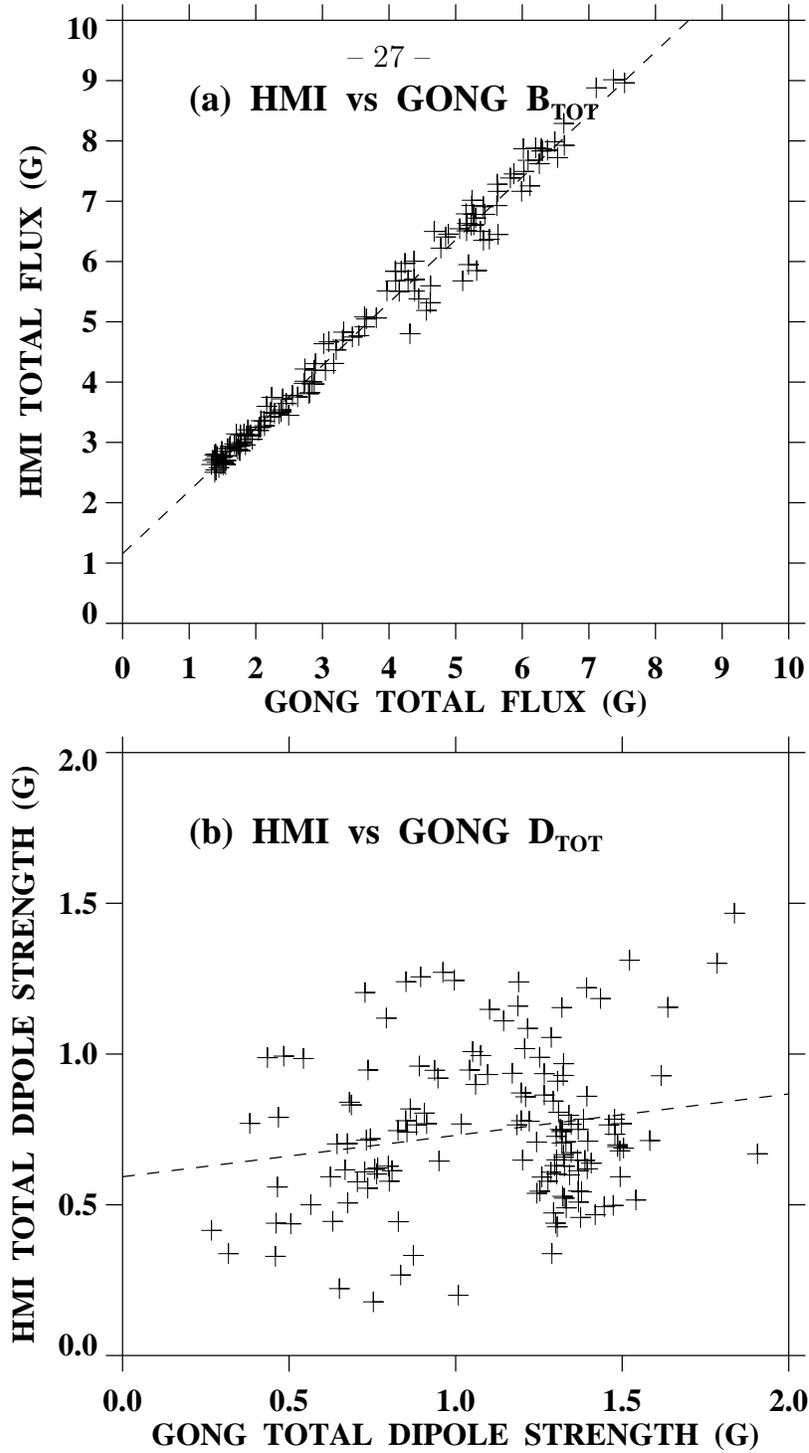}}
\vspace{-1.7cm}
\caption{Scatter plots of (a) $B_{\rm tot}$(HMI) against $B_{\rm tot}$(GONG), 
and (b) $D_{\rm tot}$(HMI) against $D_{\rm tot}$(GONG).  Each cross 
represents a CR.  Although the HMI and GONG total photospheric fluxes 
are in good agreement with each other (aside from a zero-point offset), 
their total dipole strengths are poorly correlated.}
\end{figure}

\clearpage
\begin{figure}
\vspace{-4.3cm}
\centerline{\includegraphics[width=42pc]{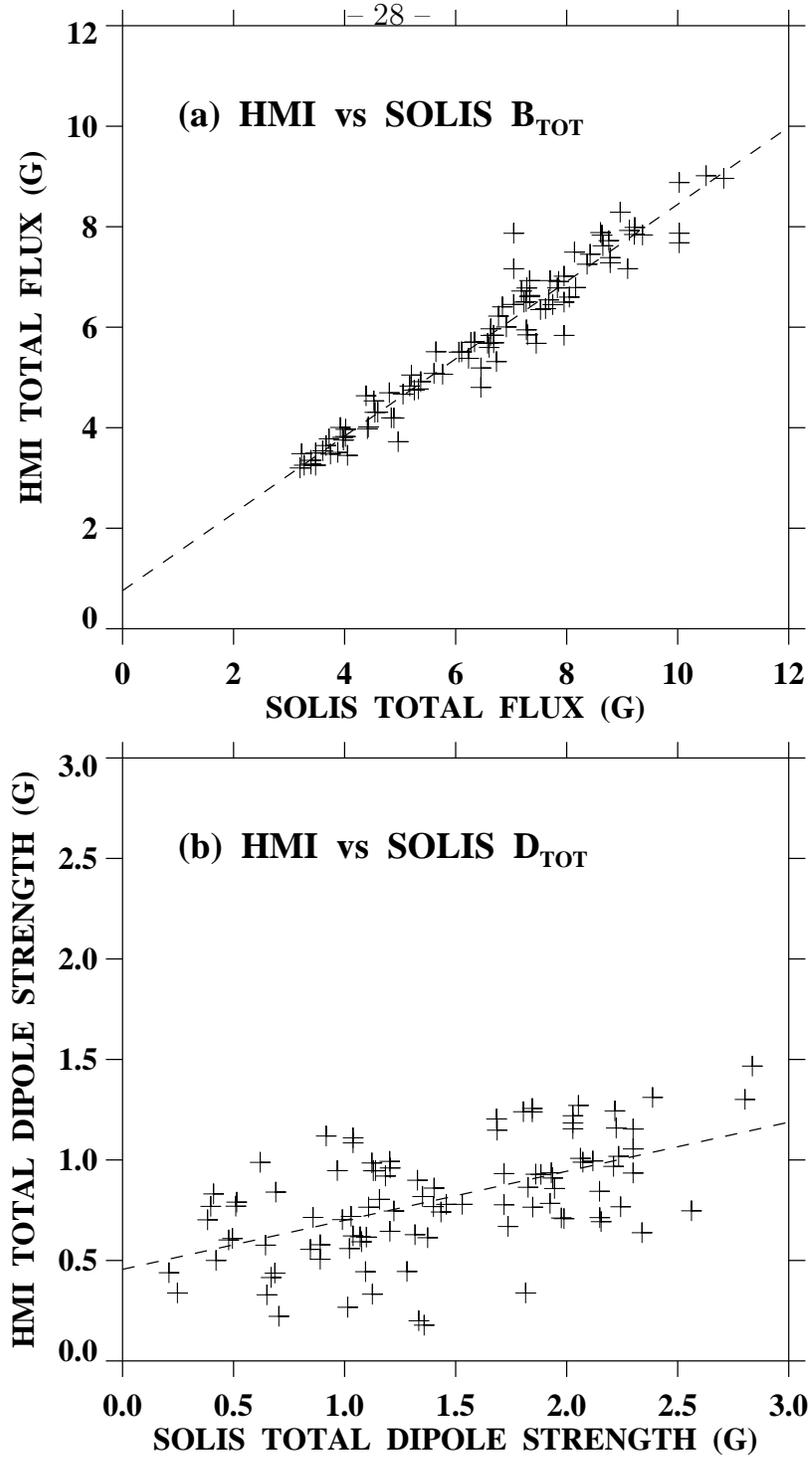}}
\vspace{-1.7cm}
\caption{Scatter plots of (a) $B_{\rm tot}$(HMI) against $B_{\rm tot}$(SOLIS), 
and (b) $D_{\rm tot}$(HMI) against $D_{\rm tot}$(SOLIS).  The SOLIS total 
fluxes are just 1.3 times larger than the HMI total fluxes, but their 
dipole strengths are 4.1 times larger than the corresponding HMI values.}
\end{figure}

\clearpage
\begin{figure}
\vspace{-3.8cm}
\centerline{\includegraphics[width=42pc]{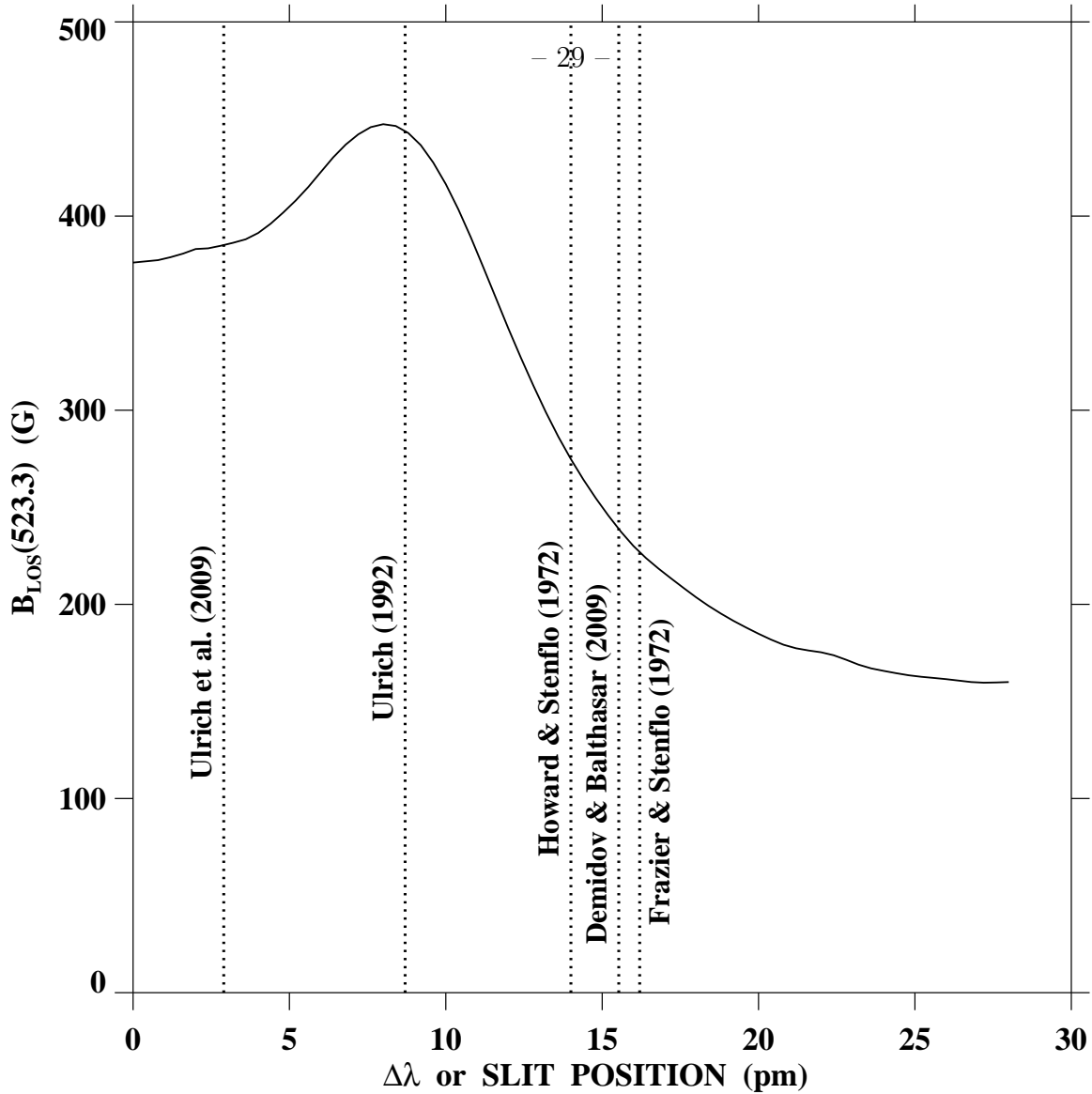}}
\vspace{-5.5cm}
\caption{Derived field strength $B_{\rm los}(523.3)$ as a function of 
$\Delta\lambda$, the wavelength position on the Fe I 523.3 nm line profile 
where the Zeeman shift is measured (using the line bisector method).  
The curve is reproduced from Figure 7 of Ulrich et al. (2009), and 
is based on left- and right-circularly polarized line profiles of a 
plage region observed with the MWO magnetograph on 2007 July 13.  
The vertical dotted lines mark the center positions of the exit slits 
used in the 523.3 nm measurements of Howard \& Stenflo (1972), 
Frazier \& Stenflo (1972), Ulrich (1992), Ulrich et al. (2009), and 
Demidov \& Balthasar (2009).  The curve peaks at $\Delta\lambda\sim 8$ pm, 
close to the slit position of Ulrich (1992).  Howard \& Stenflo, 
Frazier \& Stenflo, and Demidov \& Balthasar obtained lower values 
of $\delta^{-1} = B_{\rm los}(523.3)/B_{\rm los}(525.0)$ because 
they placed their slits farther out in the 523.3 nm line wings 
($\Delta\lambda\sim 15$ pm), where $B_{\rm los}(523.3)$ falls by a 
factor of order 2.  The slits had total widths of 
16 pm (Howard \& Stenflo 1972), 17.5 pm (Frazier \& Stenflo 1972), 
5.0 pm (Ulrich 1992; Ulrich et al. 2009), and 24.84 pm (Demidov \& 
Balthasar 2009).}
\end{figure}

\clearpage
\begin{figure}
\vspace{-2.0cm}
\centerline{\includegraphics[width=35pc]{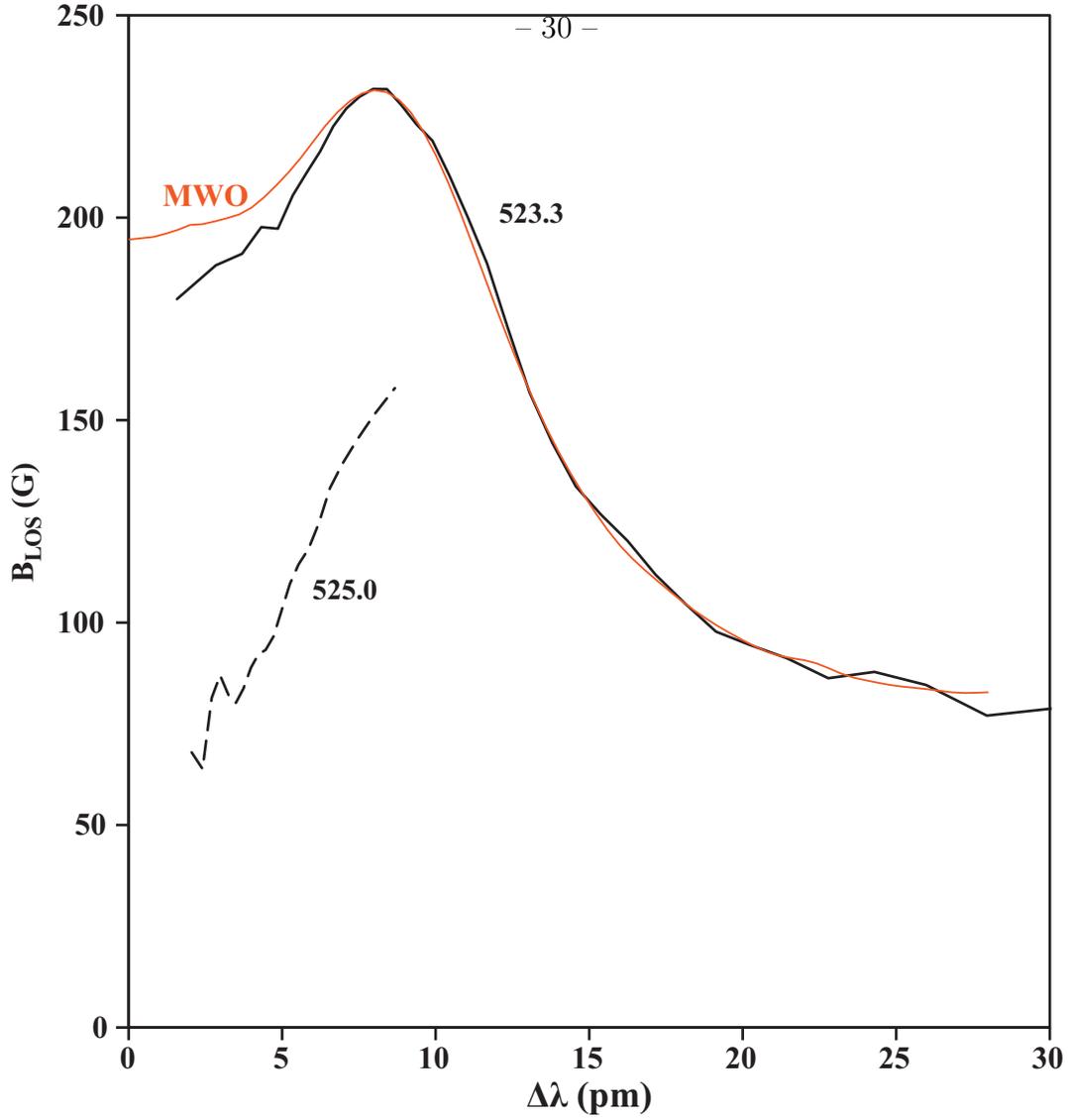}}
\vspace{-1.0cm}
\caption{Field strengths $B_{\rm los}$ derived as a function of 
wavelength position $\Delta\lambda$ from NSO/FTS measurements of the 
Fe I 523.3 nm (solid black curve) and Fe I 525.0 nm (dashed black curve) 
line profiles.  For comparison, the $B_{\rm los}(523.3)$ curve of Figure 7 
has been scaled downward by a factor of $\sim$2 and replotted in red.  
The line bisector method was applied to FTS spectra in left- and 
right-circularly polarized light from a spotless plage region 
observed on 1979 April 30 at $\cos\rho = 0.91$.  Note the strong and 
opposite variations of $B_{\rm los}(525.0)$ and $B_{\rm los}(523.3)$ 
in the line wings, and the remarkable similarity between the shapes 
of the FTS and MWO $B_{\rm los}(523.3)$ curves, despite having been 
derived using completely different instruments and observing techniques.  
Ulrich (1992), Howard \& Stenflo (1972), Frazier \& Stenflo (1972), 
and Demidov \& Balthasar (2009) centered their 525.0 nm exit slits 
at 3.9, 4.65, 5.5, and 6.21 pm, respectively.  The much larger values 
of $\delta^{-1}$ obtained by Ulrich may be a consequence of adopting 
relatively small values of $\Delta\lambda$ for both his 523.3 and 
525.0 nm measurements.}
\end{figure}

\clearpage
\begin{figure}
\vspace{-3.3cm}
\centerline{\includegraphics[width=40pc]{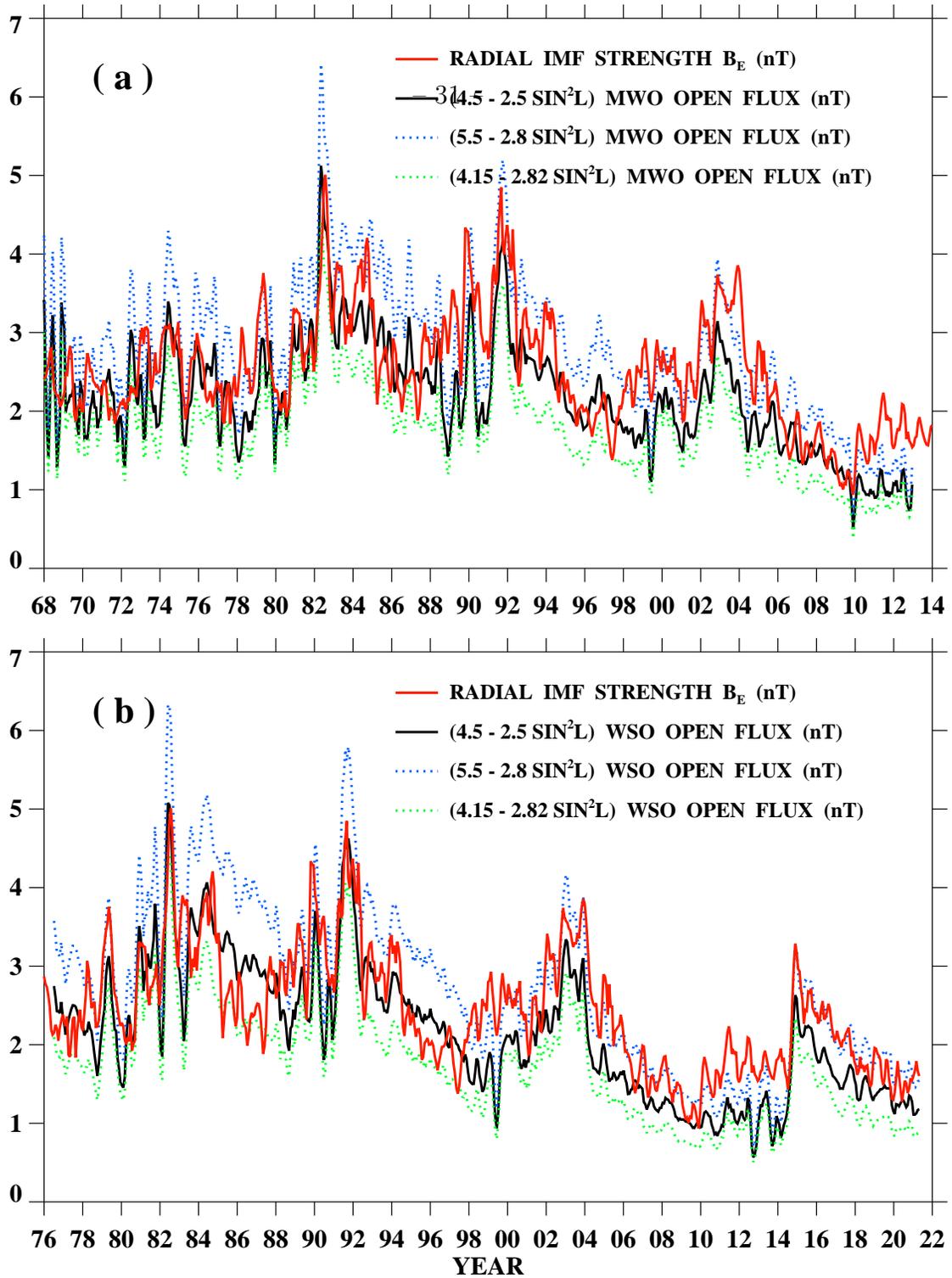}}
\vspace{-1.2cm}
\caption{Comparison between the observed radial IMF strength during 
1968--2021 and the total open fluxes derived from the (a) MWO and (b) WSO 
photospheric field maps, after applying three different versions of 
the Fe I 525.0 nm saturation correction: $\delta^{-1} = (4.5 - 2.5\sin^2L)$ 
(Ulrich 1992); $\delta^{-1} = (5.5 - 2.8\sin^2L)$ (Ulrich et al. 2009); 
and $\delta^{-1} = (4.15 - 2.82\sin^2L)$ (Ulrich et al. 2009).  As in 
Figure 1, $R_{\rm ss} = 2.5$ $R_\odot$ and all curves represent 3-CR 
running averages.  The best overall fit is obtained using the 
$\delta^{-1} = (4.5 - 2.5\sin^2L)$ scaling factor.}
\end{figure}

\clearpage
\begin{figure}
\vspace{-3.3cm}
\centerline{\includegraphics[width=40pc]{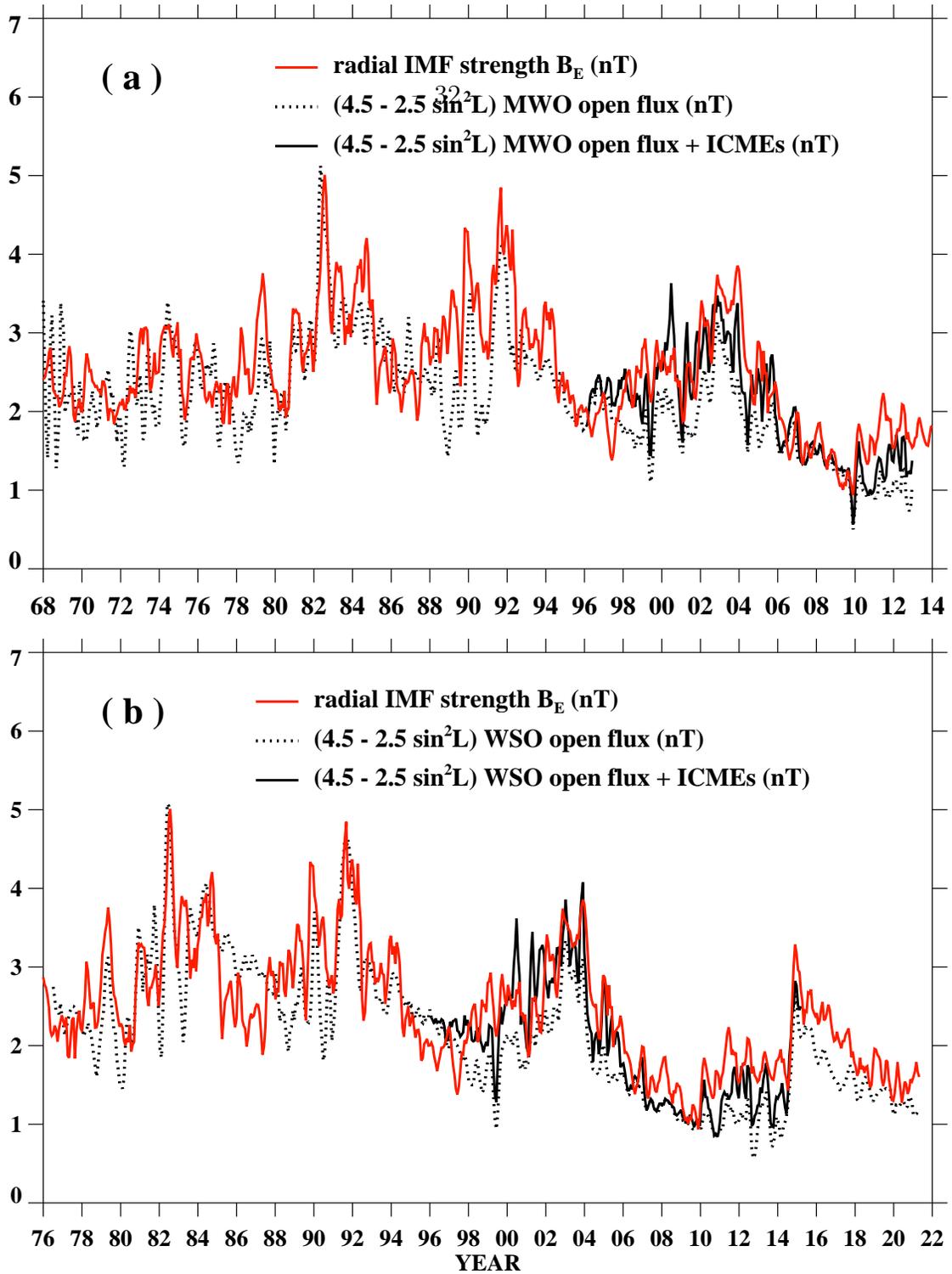}}
\vspace{-1.2cm}
\caption{Effect of adding the contribution of ICMEs to the (a) MWO and (b) WSO 
total open fluxes, corrected using the $\delta^{-1} = (4.5 - 2.5\sin^2L)$ 
scaling factor.  Near-Earth ICMEs during 1996--2015 were identified 
using the online Richardson--Cane catalog and assigned radial 
field strengths from the OMNIWeb database (see Wang \& Sheeley 2015).  
ICMEs contributed $\sim$23\% of the interplanetary flux during 1999--2002 
and $\sim$18\% during 2011--2014; their inclusion improves the agreement 
between the predicted and observed IMF strength during the rising 
and maximum phases of the solar cycle.}
\end{figure}

\end{document}